\documentclass[a4paper, oneside, twocolumn, notitlepage, 10pt]{extarticle_ecoc}
\usepackage{ecoc}
\usepackage{csquotes}
\usepackage{siunitx}
\DeclareSIUnit\count{count}
\DeclareSIUnit\bit{b}
\DeclareSIUnit\bitl{bit}

\begin{document}
\selectlanguage{american}    % Standard Language

%-------------------------------------------------- Title -----------------------------------------------------%

\title{Characterization of Electro-Optical Devices with Low Jitter Single Photon Detectors -- Towards an Optical Sampling Oscilloscope Beyond \SI{100}{\giga\hertz}}

%------------------------------------------------- Authors-----------------------------------------------------%

\author{
  % Helmut Fedder
  % Steffen Österwind
  % Markus Wick
  % Fabian Ollbrich
  % Peter Michler
  % Thomas Veigel
  % Manfred Berroth
  % Michael Schlagmüller
  Helmut Fedder\textsuperscript{(1)}, Steffen Oesterwind\textsuperscript{(2)}, Markus Wick\textsuperscript{(3)},
  Igor Shavrin\textsuperscript{(1)}, Michael Schlagm{\"u}ller\textsuperscript{(1)},\\
  Fabian Olbrich\textsuperscript{(4)}, Peter Michler\textsuperscript{(4)}, Thomas Veigel\textsuperscript{(5)}, Manfred Berroth\textsuperscript{(5)},\\
  Nicolai Walter\textsuperscript{(6)}, Wladick Hartmann\textsuperscript{(6)}, Wolfram Pernice\textsuperscript{(6)}, Vadim Kovalyuk\textsuperscript{(7)},
}

%Helmut Fedder and Michael Schlagm{\"u}ller, 

\maketitle                  % Create title and author

%------------------------------------------ Description of Authors ----------------------------------------------%

\begin{strip}
 \begin{author_descr}

  %  \textsuperscript{(1)} Swabian Instruments, Frankenstr. 39, 71701 Schwieberdingen, \textcolor{blue}{\uline{helmut@swabianinstruments.com}}
  %  \textsuperscript{(2)} Institut f{\"u}r elektrische und optische Nachrichtentechnik, Pfaffenwaldring 47, 70569 Stuttgart,
  %  \textsuperscript{(3)} Institut f{\"u}r Intelligente Sensorik und Theoretische Elektrotechnik, Pfaffenwaldring 47, 70569 Stuttgart,
  %  \textsuperscript{(4)} Institut f{\"u}r Halbleiteroptik und Funktionelle Grenzfl{\"a}chen, Allmandring 3, 70569 Stuttgart

   {\small \textsuperscript{(1)} Swabian Instruments, Frankenstr. 39, 71701 Schwieberdingen, Germany, \textcolor{blue}{\uline{helmut@swabianinstruments.com}}}

   {\small \textsuperscript{(2)} Selfnet e.V., Allmandring 8a, 70569 Stuttgart, Germany}

   {\small \textsuperscript{(3)} Institut f{\"u}r Halbleiteroptik und Funktionelle Grenzfl{\"a}chen, Allmandring 3, 70569 Stuttgart, Germany}

   {\small \textsuperscript{(4)} Institut f{\"u}r Intelligente Sensorik und Theoretische Elektrotechnik, Pfaffenwaldring 47, 70569 Stuttgart, Germany}
   
   {\small \textsuperscript{(5)} Institut f{\"u}r Elektrische und Optische Nachrichtentechnik, Pfaffenwaldring 47, 70569 Stuttgart, Germany}

   {\small \textsuperscript{(6)} Westf{\"a}lische Wilhelms-Universit{\"a}t M{\"u}nster, Center for Nanotechnology, Heisenbergstra{\ss}e 11, 48149 M{\"u}nster}

   {\small \textsuperscript{(7)} Scontel, 5/22 Rossolimo Str., 119021 Moscow, Russia}

 \end{author_descr}
\end{strip}

%\setstretch{1.1}

%-------------------------------------------------- Abstract ---------------------------------------------------------%

\begin{strip}
  \begin{ecoc_abstract}
    We showcase an optical random sampling scope that exploits single photon counting
    and apply it to characterize % 18
    optical transceivers. We study single photon detectors with
    a jitter down to \SI{40}{\pico\second}. The method can be extended beyond \SI{100}{\giga\hertz}.
  \end{ecoc_abstract}
\end{strip}

%-------------------------------------------------- Introduction Section -------------------------------------------------------%

\section{Introduction}
Single photon counting \cite{Becker05} has found broad applications within quantum technologies, such as quantum sensing \cite{Degen17},
quantum information \cite{Bennett00} and quantum communication \cite{Gisin07}. More recently it was proposed \cite{Korz18} that
low jitter single photon detectors could be used to realize optical random sampling scopes with a bandwidth
well beyond \SI{100}{\giga\hertz}, a range not accessible with existing measurement instrumentation.
Such instrumentation is suitable to support research on next generation electro-optical devices such as EOMs and \mbox{VCSELs}.
In this demo session paper, we showcase a proof of principle measurement and characterize
commercial SFP\raisebox{0.5ex}{+} modules using both Single Photon Avalanche Detectors (SPADs) and
Superconducting Nanowire Single Photon Detectors (SNSPDs) with down to \SI{40}{\pico\second} jitter (FWHM).
%in conjunction with a low jitter multichannel time-to-digital-converter implemented
%in a \mbox{Kintex 7} \mbox{FPGA}.
Both detectors are capable of detecting signals in the femtowatt regime.
The SPAD provides an effective measurement bandwidth of about \SI{11}{\giga\hertz}.
The SNSPD provides a measurement that is free of
typical detector artefacts, such as ringing and non-linearity.
With the jitter of current SNSPDs reaching into the few picosecond regime \cite{Korz18} the method holds promise
to enable a measurement bandwidth well beyond \SI{100}{\giga\hertz}.
This research is timely and innovative and will appeal to the audience.
All data shown below will be measured live at the demo session setup.
Visitors will have access to a modern graphical touch interface
that will enable them to control all measurement parameters
and experience the single photon optical sampling technique on site.
\section{Measurement Setup}
\begin{figure}[t]
  \centering
  \makebox[0pt][l]{\raisebox{58mm}{a)}}\includegraphics[width=78mm]{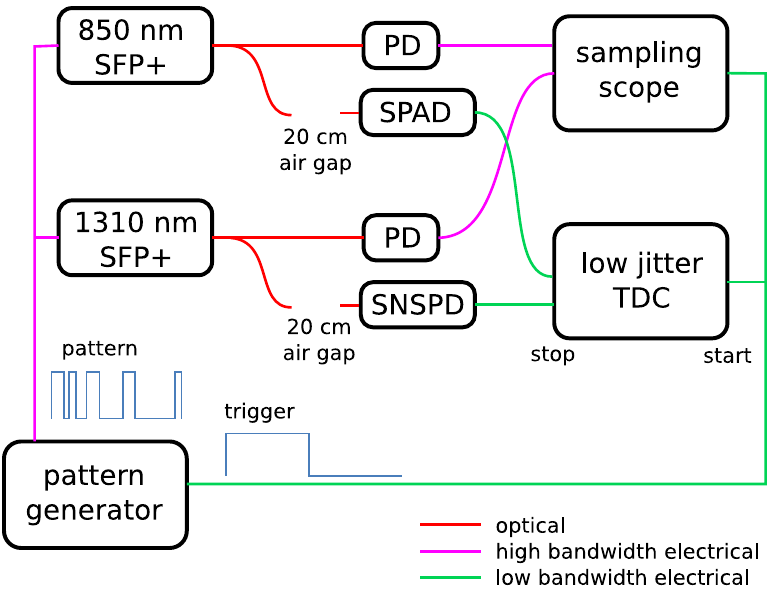}\\
  \makebox[0pt][l]{\raisebox{42mm}{b)}}\includegraphics[width=76mm]{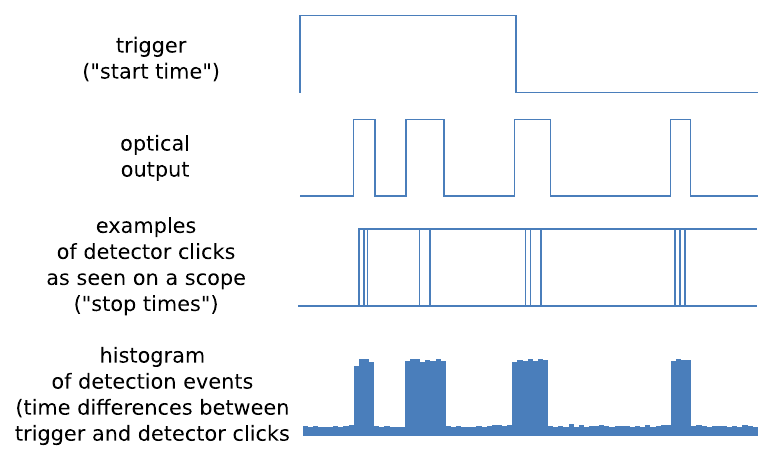}
  \caption{a) optical and electrical setup, b) measurement principle.}
   \label{fig:setup}
\end{figure}
Fig.\ref{fig:setup} illustrates the measurement setup as presented in the
ECOC demo session. The setup measures the modulation response and eye diagram of an \SI{850}{\nano\meter} MMF (FINISAR FTLX8574D3BCL, 10\,Gbit/s)
and a \SI{1310}{\nano\meter} SMF (Juniper EX-SFP-10GE-LR-C-UL, 10\,Gbit/s) SFP\raisebox{0.5ex}{+} module, simultaneously with the single photon counting approach exploited
herein and with a state-of-the-art test system (\SI{30}{\giga\hertz} InGaAs photodiode, Thorlabs DXM30BF, and \SI{50}{\giga\hertz} oscilloscope,
Keysight Infiniium). Each optical output of the SFP\raisebox{0.5ex}{+} modules is split with fiber splitters. One arm of each
output is directed to a photodiode (PD) and oscilloscope. The second arm is left open and the light is allowed
to travel through about \SI{20}{\centi\meter} air gap after which a fraction of the intensity is picked up with an optical fiber again
and directed to the SPAD (Silicon APD, IDQ 100) and SNSPD (Scontel) for \SI{850}{\nano\meter} and \SI{1310}{\nano\meter}, respectively.
The air gap provides an attenuation on the order of 100\,dB and the distance and coupling efficiency
of the gap is used to adjust the photon count rates on the detectors under CW
illumination to about 1\,Mcounts/s. This count rate ensures that
the average time between two photon detection events is much larger than the detector dead times
(about \SI{50}{\nano\second}) such that the single photon detectors are operated
in the linear range well below saturation.
The outputs from the single photon counters are connected to two independent channels
of a multichannel time-to-digital converter (TDC),
%(Swabian Instruments, Time Tagger Ultra 8)
that is implemented in a \mbox{Kintex 7} FPGA
by exploiting carry chains of the dedicated adders as tapped delay lines
\cite{Favi09}. A low jitter pattern generator delivers two synchronous output patterns:
one output generates the 10\,Gbit/s pattern that is applied to the SFP\raisebox{0.5ex}{+} modules. A second output generates
a trigger pulse for each bit pattern that is applied to the oscilloscope and to a third channel of the TDC.
The random sampling measurements proceeds as follows (see Fig.~\ref{fig:setup}b). The optical pattern and the trigger pulses are
generated repeatedly with a rate on the order of 10\,MHz. The triggers are used as "start" clicks and
detected photons are used as "stop" clicks. Note that with the given photon count rates (imposed by the detector dead times),
on average less than one photon is detected per cycle. The TDC
measures the time differences between "start" clicks and "stop" clicks and accumulates them in a histogram,
thereby providing a random sampled representation of the optical intensity.
It is instructive to consider the analog bandwidth of distinct signal paths of the system (see also Fig.~\ref{fig:setup}a).
The photodiodes are connected to the oscilloscope via suitable 2.92 mm coaxial lines that provide \SI{40}{\giga\hertz} analog bandwidth.
By contrast, the analog bandwidth of the electrical pulses that are output by
the single photon detectors is only a few GHz. Thus, simple RG\,316 SMA cables
and input discriminators with moderate bandwidth (Texas Instruments LMH7322, maximum toggle rate \SI{3.9}{\giga\hertz}, jitter \SI{572}{\femto\second}) are enough
to deliver the detector signals to the TDC. This aspect is important: in an optical sampling
scope based on single photon detectors, the measurement bandwidth is determined by the \textit{jitter} of the
single photon detectors and \textit{not} by the analog bandwidth of the receiver.
This enables us to conceive an optical sampling scope with a bandwidth that could reach into the THz regime
without the need of handling high frequency electrical signals.

\section{Results}
\begin{figure}[t]
  \centering
  \includegraphics[width=78mm]{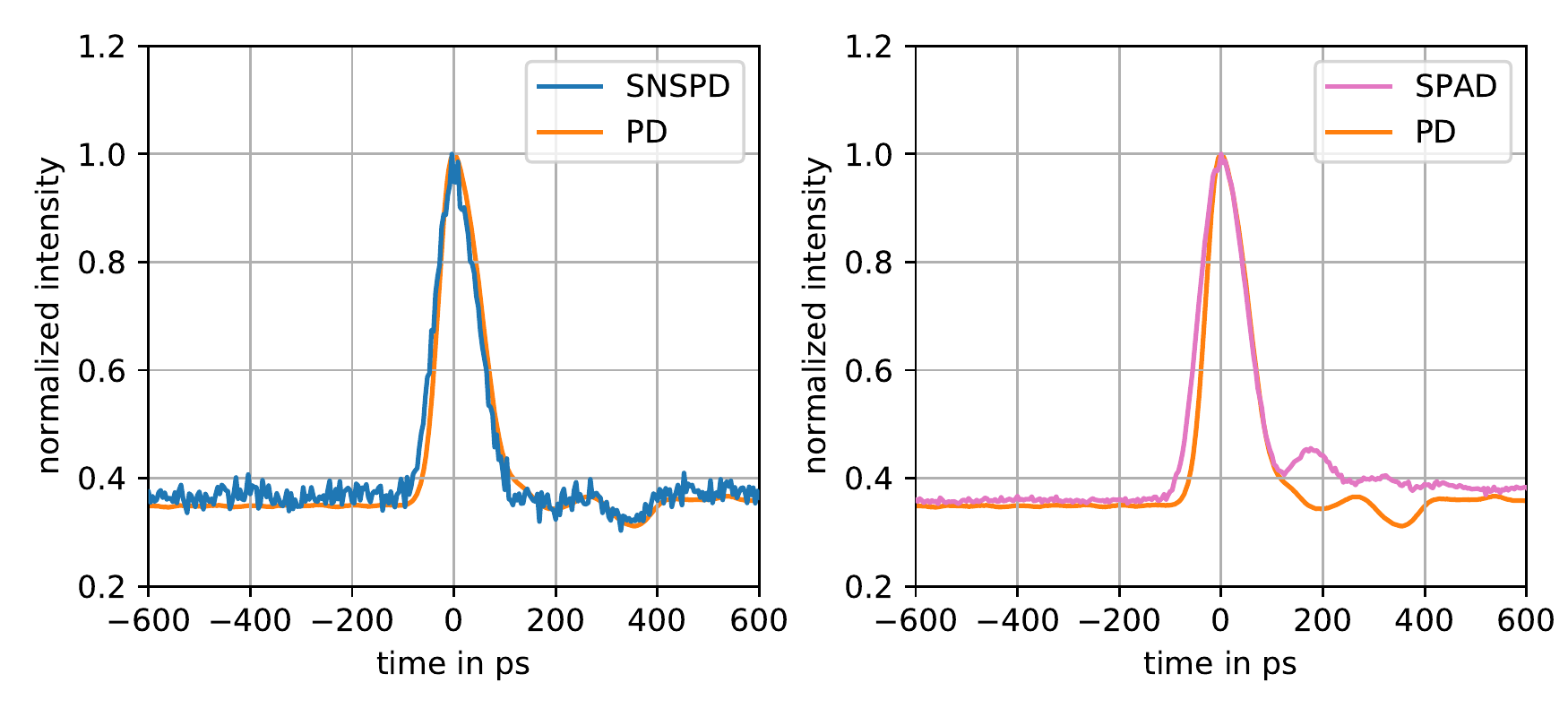}\makebox[0pt][l]{\raisebox{32mm}{\hspace{-78mm}a)}}\makebox[0pt][l]{\raisebox{32mm}{\hspace{-39.5mm}b)}}
  \caption{Optical pulse response for isolated 1 bit pulses (duration \SI{100}{\pico\second}).
  Single photon counting results (orange) are compared to the SNSPD (blue) and SPAD (pink).}
   \label{fig:pulse}
\end{figure}

We first characterize the pulse response of both single photon detectors by applying isolated 1 bit pulses to
the SFP\raisebox{0.5ex}{+} modules, providing nominally \SI{100}{\pico\second} long optical output pulses. Fig.~\ref{fig:pulse} shows the measured optical pulses.
The results obtained with single photon counting are compared to the measurements with the photodiodes
and they are generally in good agreement.
At the same time, distinct measurement artefacts are apparent. In case of the photo diode
the main peak is slightly skewed and weak ringing is observed, seen as a shoulder at about +\SI{100}{\pico\second}.
The single photon counting measurements provide an ideal Gaussian pulse shape and are free of ringing and electrical reflections.
On the other hand, the SPAD is subject to afterpulsing\cite{Ziarkash17}, which unfolds as a weak exponential tail
with a time constant of about \SI{2}{\nano\second}, here modulated with weak oscillations from the VCSEL itself.
The response function of the SNSPD is artefact free: an ideal symmetric pulse response is obtained that is only limited by the
Gaussian broadening induced by the detector jitter. Please note that the weak dip at about +\SI{270}{\pico\second} is a true intensity dip
that is present in the optical signal.

\begin{figure}[t]
  \centering
  \includegraphics[width=80mm]{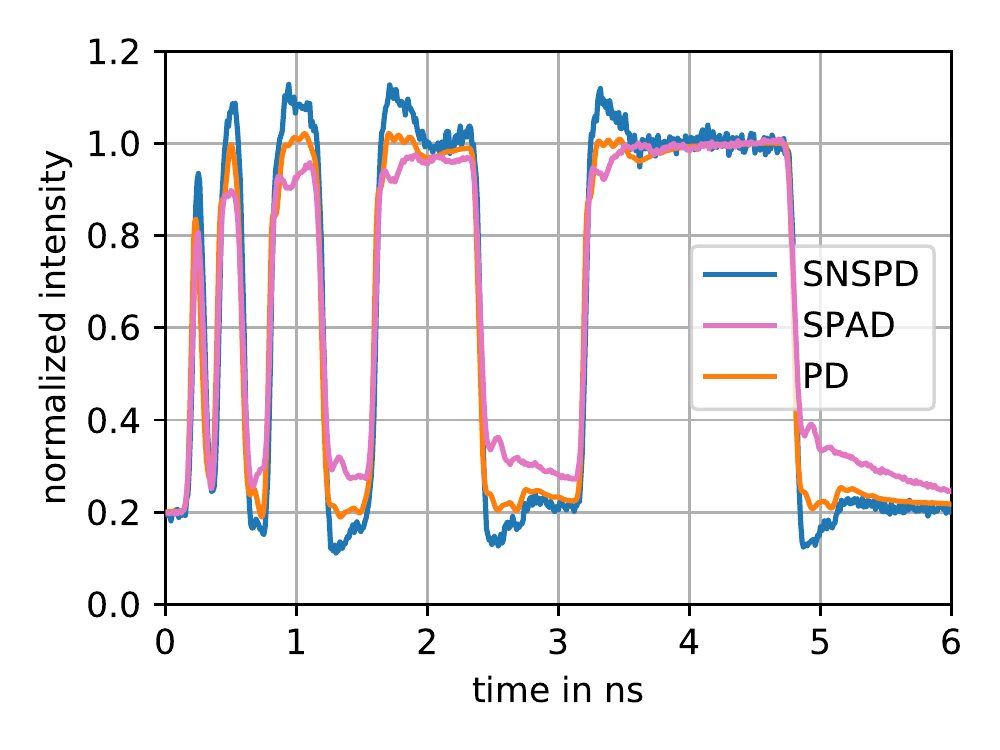}
  \caption{Pulse patterns. The nominal pulse widths are 100, 200, 400, 800 and 1600 ps (from left to right).}
   \label{fig:pattern}
\end{figure}

\begin{figure}[t]
  \centering
  \includegraphics[width=37mm]{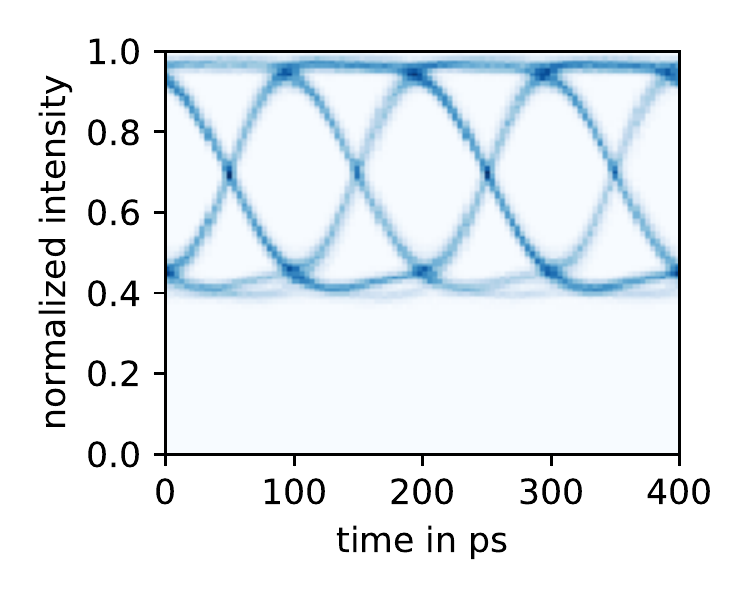}\makebox[0pt][l]{\raisebox{9mm}{\hspace{-23mm}\fbox{\tiny{SPAD, \SI{850}{\nano\meter}}}}}
  \includegraphics[width=37mm]{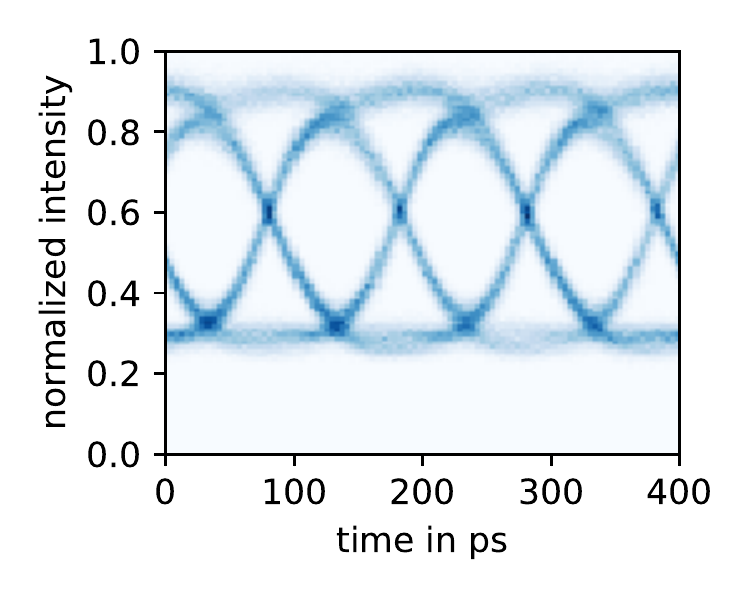}\makebox[0pt][l]{\raisebox{9mm}{\hspace{-24mm}\fbox{\tiny{SNSPD, \SI{1310}{\nano\meter}}}}}\\
  \includegraphics[width=37mm]{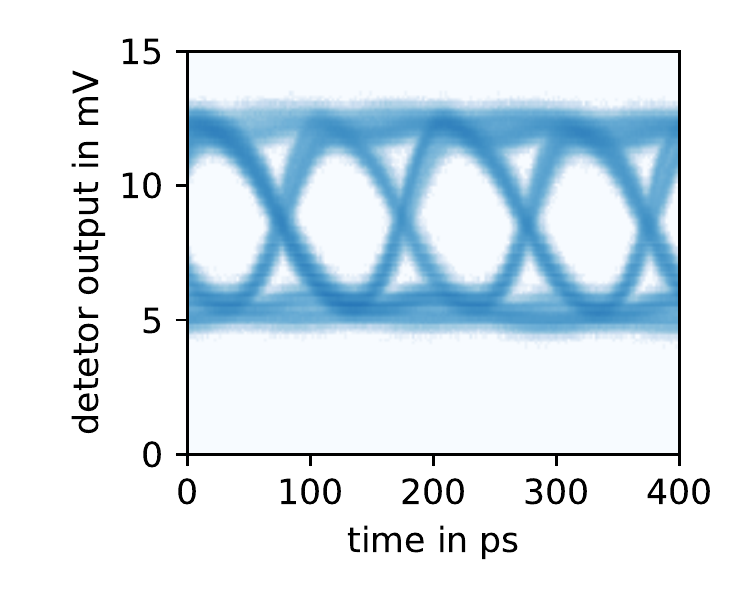}\makebox[0pt][l]{\raisebox{9mm}{\hspace{-21mm}\fbox{\tiny{PD, \SI{850}{\nano\meter}}}}}
  \includegraphics[width=37mm]{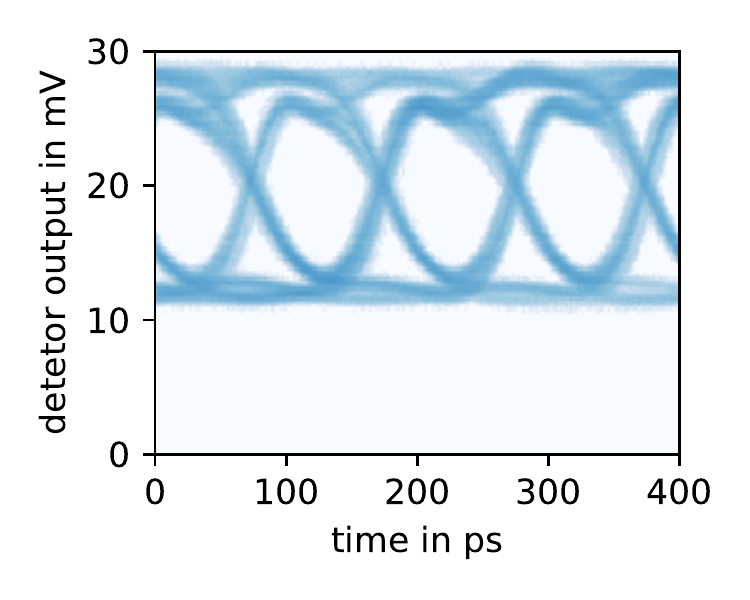}\makebox[0pt][l]{\raisebox{9mm}{\hspace{-22mm}\fbox{\tiny{PD, \SI{1310}{\nano\meter}}}}}
  \caption{Eye diagrams.}
   \label{fig:eye}
\end{figure}

To better illustrate the effects of the afterpulsing present with the SPAD,
we proceed to measure optical square pulses with increasing period (see Fig.~\ref{fig:pattern}).
The afterpulsing tails are now more apparent. Since the detector pulse
response enters the measurement via a convolution, for
pulses with longer period, the afterpulsing tail, respectively ringing is accumulated over a longer range
and distorts the signal more strongly. For the SNSPD, an artefact free measurement
is obtained. Note that the noise on the single photon counting measurements are purely due to
shot noise and can be improved to the desired precision by averaging for a longer time.
For the present proof of principle demonstration, we proceed and measure the eye diagrams of the
SFP\raisebox{0.5ex}{+} modules (see Fig.~\ref{fig:eye}). For this measurement, we apply random 8b10b symbols
to the SFP\raisebox{0.5ex}{+} modules. To acquire this data with single photon counting,
one histogram is accumulated for each 8b10b symbol and the data is
combined to a two-dimensional histogram.
With this measurement, the effect of the skewed pulse response
of the photodiode becomes apparent. It unfolds as a distortion of the respective
eye diagrams. By contrast, the single photon counting measurements are free of distortions due to
their inherent Gaussian pulse responses.

% ToDo: change to absolute counts, figures a)b)c)
% normalized ?

\section{Conclusions}
We have demonstrated a proof-of-principle optical sampling scope based on low jitter single photon detectors
and have applied it to measure the pulse response and eye diagram of two commercial SFP\raisebox{0.5ex}{+} modules.
Thereby, we have studied the performance of two types of modern single photon counting detectors,
one, a Si-SPAD, that is operated at room temperature and useable in the visible wavelength range,
and a second one, an SNSPD, that requires cryogenic operation and that is optimized for the \SI{1310}{\nano\meter}
and \SI{1550}{\nano\meter} telecom bands 
and useable over an extended optical wavelength range.
We observed that both detectors provide accurate and undistorted eye diagram measurements, which we explained
with their nearly ideal Gaussian pulse responses. We found that the SPAD results in significant distortion when
measuring a step response, which we attributed to the afterpulsing tail. The SNSPD, which does not suffer
from afterpulsing, provided an ideal step response measurement. The measurement scheme
allows for the detection of very low optical powers down to the femtowatt regime and
can provide very high optical measurement bandwidth without requiring high
bandwidth electrical signals. The measurements presented herein are limited by the jitter of the
photodetectors, which is \SI{40}{\pico\second} (FWHM) for the SPAD and \SI{50}{\pico\second} (FWHM) for the SNSPD, respectively. This translates
into an effective measurement bandwidth of \SI{11}{\giga\hertz}, respectively \SI{9}{\giga\hertz}.
Recently SNSPDs with \SI{2.7}{\pico\second} jitter
(FWHM) were reported\cite{Korz18}, translating into an effective measurement bandwidth of about \SI{160}{\giga\hertz}.
At present, there is no technical limit known for the jitter achievable with single photon detectors
and it is expected that it will be significantly further reduced within the
next few years. This opens
the perspective for optical sampling scopes with a bandwidth well beyond \SI{100}{\giga\hertz}.
The TDC used in the present study has an RMS jitter of \SI{8}{\pico\second} and will be the limiting factor
in future optical sampling scopes. This calls for the implementation
of low jitter multichannel TDCs with short dead times (no more than few ns)
that will enable a novel class of single photon based optical sampling scopes. 

\section{Acknowledgements}
We thank Matthew Shaw and Boris Korzh from JPL and Vikas Anant from Photon Spot for
fruitful discussions.

%-------------------------------------------------- Bibliography Section -------------------------------------------------------%

\bibliographystyle{abbrv}
\begin{spacing}{1.35}

\end{spacing}
\vspace{-4mm}

%%%%%%%%%%%%%%%%%%%%%%%%%%%%%%%%%%%%%%%%%%%%%
%---------------------------------------------- End of Document -----------------------------------------------%
\end{document}